\documentclass[12pt]{article}

\makeatletter
\def\cl@chapter{\@elt {theorem}}
\renewcommand\section{\@startsection{section}{1}{\z@}%
                                  {-3.5ex \@plus -1ex \@minus -.2ex}%
                                  {2.3ex \@plus.2ex}%
                                  {\normalfont\large\bfseries}}
\makeatother

\usepackage[backend=bibtex8,sorting=none]{biblatex}
\renewbibmacro{in:}{}
\bibliography{latticeImaginary} 

\AtEveryBibitem{\clearfield{month}}
\AtEveryBibitem{\clearfield{doi}}
\AtEveryBibitem{\clearfield{url}}
\AtEveryBibitem{\clearfield{issn}}
\AtEveryBibitem{\clearfield{number}}
\AtEveryBibitem{\clearfield{eid}}

\usepackage{graphicx} 
\graphicspath{ {./images/} }
\usepackage[a4paper,top=1.2in, left=0.3in, right=0.2in, bottom=1in]{geometry}
\usepackage[T1]{fontenc}
\usepackage{tikz}
\usetikzlibrary{intersections,positioning,shapes.misc,calc,arrows}
\usepackage{mathrsfs}
\usepackage{mathtools}
\usepackage{hyperref}
\usepackage{caption}

\hypersetup{colorlinks=true,linktocpage=true,citecolor=blue,urlcolor=cyan,linkcolor=red}
\usepackage{amsfonts}
\usepackage{microtype}
\usepackage[mathscr]{eucal}
\usepackage[affil-it]{authblk}

\usepackage[side]{footmisc}

\begin{document}

\title{\textbf{\Large{Quasi-hermitian lattices with imaginary \\ zero-range interactions}}}
\author{Frantisek Ruzicka \thanks{\texttt{ruzicfra@fjfi.cvut.cz}}}
\affil{\normalsize{Faculty of nuclear sciences and physical engineering \\
Czech technical university in Prague \\ Brehova 7, 115 19 Prague \\ Czech Republic}}
\affil{Nuclear physics institute, Czech academy of sciences \\ Hlavni 130, 250 86 Husinec-Rez \\  Czech republic }
\date{\normalsize{\today}}

\maketitle

\abstract{
We study a general class of $\mathscr{PT}$-symmetric tridiagonal Hamiltonians with purely imaginary interaction terms in the quasi-hermitian representation of quantum mechanics. Our general Hamiltonian encompasses many previously studied lattice models as special cases. We provide numerical results regarding domains of observability and exceptional points, and discuss the possibility of explicit construction of general metric operators (which in turn determine all the physical Hilbert spaces). The condition of computational simplicity for the metrics motivates the introduction of certain one-parametric special cases, which consequently admit closed-form extrapolation patterns of the low-dimensional results.}

\section{Introduction}

Quasi-hermitian quantum Hamiltonians have received considerable attention during recent years \cite{ScholzGeyerHahne,AliReview, BenderReview,BenderBoettcher, DoreyDunningTateo}. The reason of their popularity may be seen in the fact, that they might express genuine quantum observables by apparently non-hermitian (and often more computationally friendly) operators. From the mathematical viewpoint, these operators are characterized by a generalized condition of hermicity

\begin{equation}
\label{DieudonneEquation}
H^\dagger \Theta = \Theta H
\end{equation}

for some bounded nonsingular self-adjoint operator $\Theta > 0$, usually called the metric. Finding solutions of this equation (understood as an equation for $\Theta$) forms an integral part of quasi-hermitian representation of quantum mechanics. Exact solutions for quasi-hermitian Schr\"odinger operators are rarely encountered (with one of the few exceptions being discussed in \cite{KrejcirikBila, KrejcirikTwoParameters}). Apparently, a need for a more thorough examination of the solutions necessitates the reduction of the domain of applicability into finite-dimensional vector spaces, where complete solutions can be, in principle, found explicitly.

It is well known that the general solution of \autoref{DieudonneEquation} is by far not unique. While one is often satisfied with a single solution $\Theta$ demonstrating the quasi-hermicity of the corresponding $H$, the unitary non-equivalence of the resulting inner products motivates the search for general classes of metric operators for a single Hamiltonian, and in extreme cases even for a complete solution. In finite dimensions, such a solution is known to contain a definite number of free parameters, which is equal to the dimension of underlying vector space. This opens the possibility of explicit parametrization of such a metric. This approach has been pursued, among others, in \cite{ZnojilJacobi, squareWellMetric,graphMetric, discreteRobin}, and also the recent article \cite{myMyMy}, which may be seen as a direct predecessor of the present paper.

Finite-dimensional toy models, in addition to having diverse applications in quantum and solid-state physics \emph{per se} \cite{BoseHubbard, FanoAnderson, SSH, AubryAndre}, provide also vast possibilities for the description of various physical phenomena in simplified scenarios. Indeed, finite quasi-hermitian Hamiltonians have been succesfully used to model quantum phase transitions \cite{myPhaseTransitions}, quantum catastrophes \cite{myCatastrophes} or even simplified big-bang scenarios \cite{ZnojilBigBang}. The objects of interest in all these cases are the so-called exceptional points \cite{Kato,HeissEP}, which emerge inevitably on boundaries of observability domains. These applications provide also the main morivation of the present article.

This paper is divided into five sections. Section 2 is devoted to general discussion of the studied Hamiltonian as well as the numerical examination of its spectral properties and observability domains. In section 3, we begin with the attempt of solving \autoref{DieudonneEquation} in full generality using brute-force methods of computer algebra. Motivated by the obtained results, we devote section 4 to the possibility of having a simple complete solution for abitrary dimension $n$, which leads to the restriction of parameters into certain one-parametric subspaces. We examine the possibility of all but one parameters set to zero, as well as the Hamiltonian with full lattice interaction with strictly alternating interaction terms. The latter model proves exceptionally friendly in terms of pseudometric constructions, while all these models appear previously unnoticed in the literature. The final section is devoted to discussion of the discovered phenomena.

\section{The Hamiltonian}

\begin{figure}[h!]
    \centering
    \includegraphics[height=3.7cm,width=5.5cm]{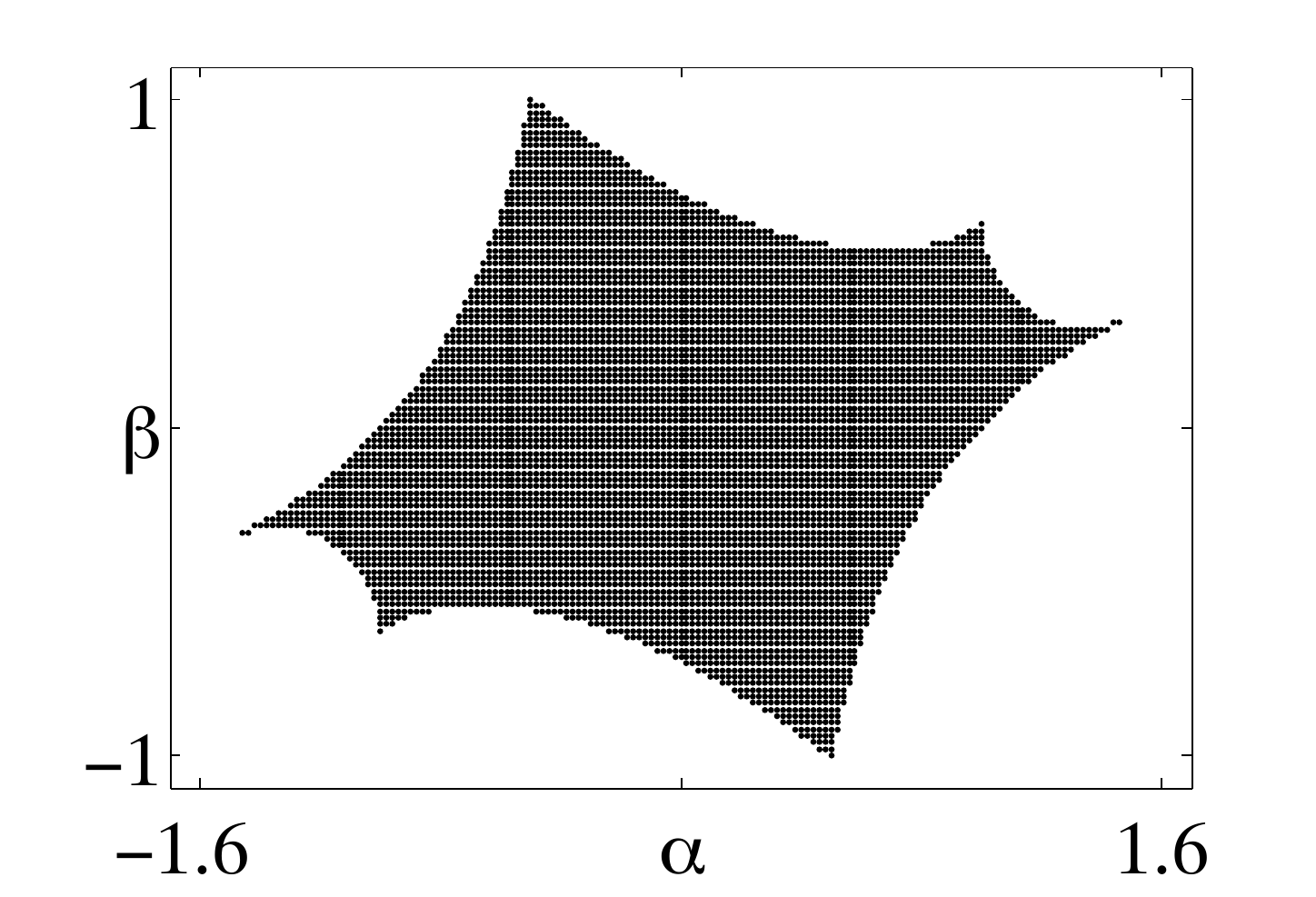} 
    \includegraphics[height=3.7cm,width=5.5cm]{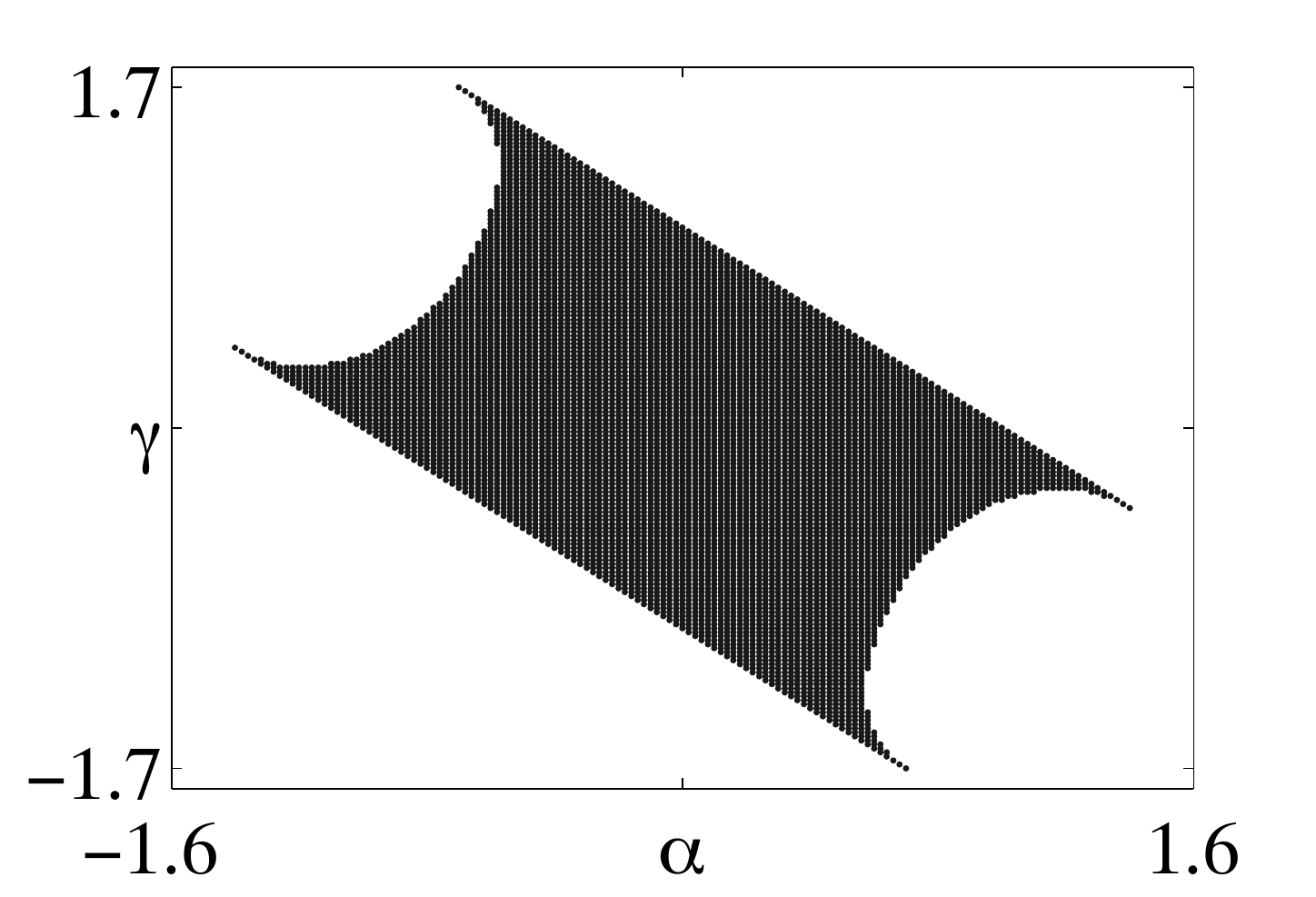} 
    \includegraphics[height=3.7cm,width=5.5cm]{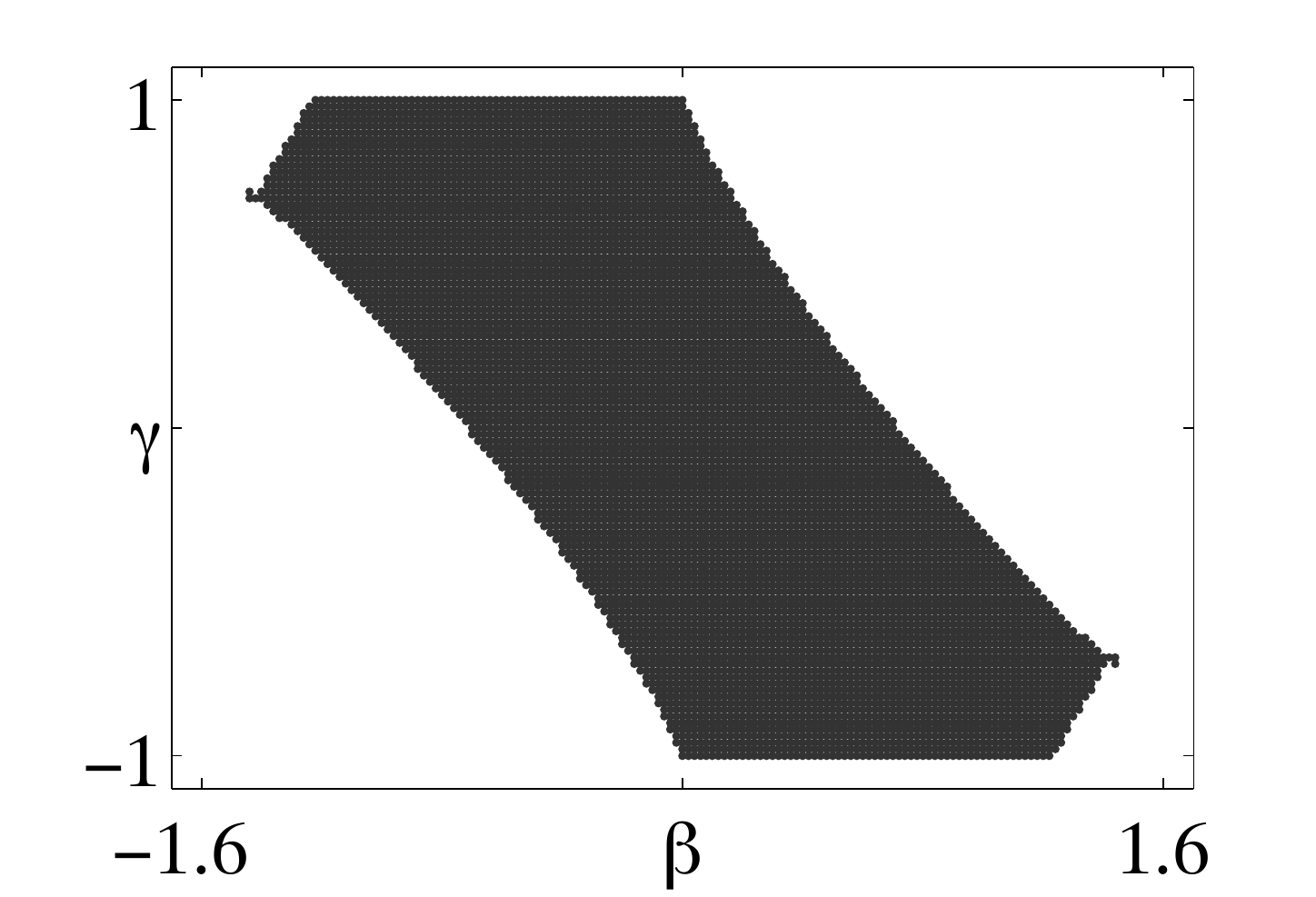} 
    \caption{\emph{Varying domains of observability for the $n=6$ model with one parameter set to zero.}}
    \label{dobs}
\end{figure}

Zero-range interaction Hamiltonians are without doubt the most studied and understood class of quantum-mechanical operators. In our finite-dimensional context, the zero-rangedness results in a family of tridiagonal matrices (see also \cite{discreteBender, ZnojilGegenbauer, discreteGraph}. These matrix Hamiltonians provide a discrete analogue for the differential Schr\"odinger operators, while at the same time having well understood spectral properties and infinite analogues acting on $\ell^2 (\mathbb{N})$ \cite{JacobiMatrices, BottcherBook}. Our Hamiltonian is intended to have a very general nature, hence the large number of parameters. The only constraint imposed is the condition of $\mathscr{PT}$-symmetry. With this restriction in mind, the most general form of our Hamiltonian is a family of $n$-parametric $2n \times 2n$ matrices

\begin{equation}
H^{(2n)} = \begin{bmatrix}
i \alpha & -1 &  &  \\
-1 & i \beta & -1 &  \\
& & \ddots &  &  \\
& & -1 & - i \beta & -1 \\
& &  & -1 & - i \alpha \end{bmatrix}
\label{ourHamiltonian}
\end{equation}

with additional parameters denoted by greek letters in alphabetical order. Although the rising number of parameters makes the model very complicated with growing dimension, the reward may be seen in the vast diversity of its spectral properties. This is demonstrated even in the next-to-trivial case $n=6$, where three domains of observability are plotted in \autoref{dobs}. The plots are made for a single parameter set to zero, in order to allow two-dimensional plotting (the stability of these patterns may be verified easily for higher dimensions)

The asymmetry of these plots serves also as an inspiration to undertake a deeper numerical experiment with single nonzero parameter. The results of this numerical experiment are summarized in \autoref{dobs1D} for 4 different next-to-boundary parameters. The domains of observability form a symmetric interval around zero, with the boundaries of observability domains being composed of exceptional points $p_{crit}$, at which the Hamiltonian ceases to be diagonalizable.

\begin{table}[h!]
\centering
\renewcommand{\arraystretch}{1}
\begin{tabular}{|c||c|c|c|c|}
  \hline
  & $n=10$ & $n=30$ & $n=50$ & $n=100$\\
  \hline \hline
   $\alpha_{crit}$ & 1.0000 & 1.0000 & 1.0000 & 1.0000 \\
  \hline
   $\beta_{crit}$ & 0.7129 & 0.7089 & 0.7082 & 0.7064 \\
  \hline
   $\gamma_{crit}$ & 0.5228 & 0.5085 & 0.5027 & 0.5015 \\
  \hline
   $\delta_{crit}$ & 0.4535 & 0.3936 & 0.3913 & 0.3828 \\
  \hline
\end{tabular}
\renewcommand{\arraystretch}{1}
    \caption{\emph{Exceptional points for $4$ different single-parametric Hamiltonians of varying dimension $n$.}}
    \label{dobs1D}
\end{table}

With exception of the parameter $\alpha$, the domains show a shrinking behavior for growing $n$. Despite this fact, the shrinking rate is decreasing quickly and numerical experiments suggest that the limit of $p_{crit}$ does not approach zero for a fixed $p$. This opens up the theoretical possibility of existence of a infinite-lattice quasi-hermitian operator in appropriatelly defined limit $n \rightarrow \infty$ for arbitrary fixed parameter and their nontrivial combinations. For finite dimensions, this indicates the existence of nontrivial regions for any possible choice of parameters (this is one of the reasons for introducing the condition of $\mathscr{PT}$-symmetry).

\section{The pseudometrics}

In this section, we address the possibility of constructing a complete solution of \autoref{DieudonneEquation} for our general Hamiltonian. We proceed in a way common in the literature and start with the construction of the so-called pseudometrics. The pseudometrics are simply metrics with the positivity condition being relaxed. In other words, they are arbitrary bounded nonsingular self-adjoint operators satisfying the compactibility condition. In a finite-dimensional Hilbert space of dimension $n$, it is known that each quasi-hermitian operator admits precisely $n$ linearly independent pseudometrics. Consequently, we are searching for a linearly independent set $\left\{ P^k_n \mid k = 1, \dots, n \right\}$ satisfying

\begin{equation}
H^\dagger_n P^k_n = P^k_n H_n^{}
\end{equation}

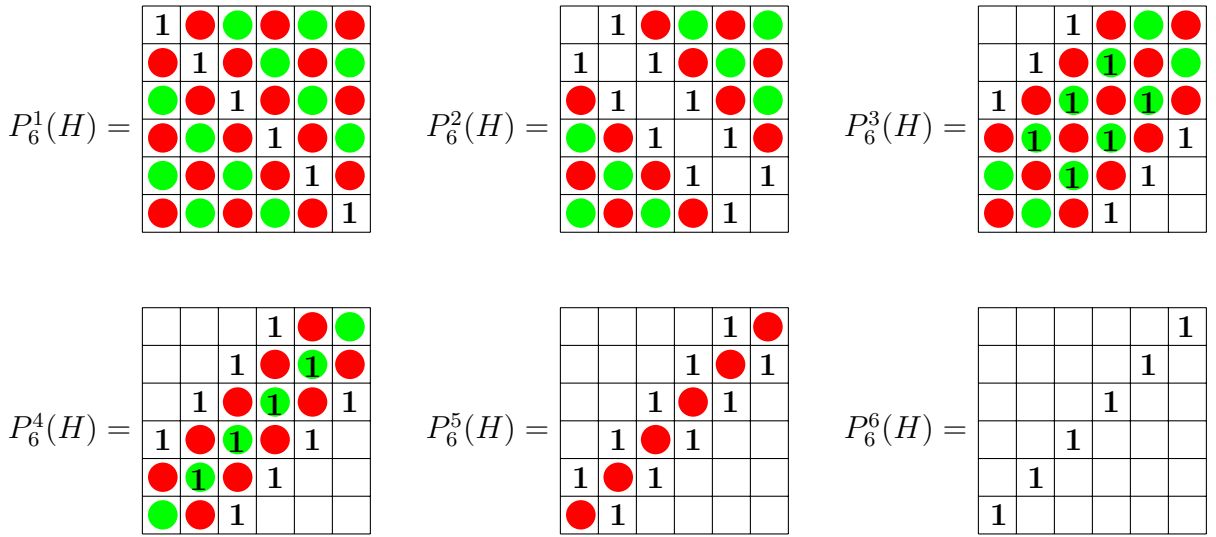
\begin{figure}[h!]
\centering
\begin{tikzpicture}[circ/.style = {circle,radius=2mm,fill=green},circ2/.style = {circle,radius=2mm,fill=red},scale=0.5]
\draw[shift={(0.5,0.5)}] (0,0) grid (6,6);
\node[below left] at (0.5,4) {\normalsize{$P^1_6 (H) =$}};
\matrix[row sep=-0.5mm,column sep=-0.2mm,shift={(1.75,1.75)}] {
\node[] {\textbf{1}}; & \node[circ2] {}; & \node[circ] {}; & \node[circ2] {}; &\node[circ] {};& \node[circ2] {};\\
\node[circ2] {}; &\node[] {\textbf{1}};& \node[circ2] {}; & \node[circ] {}; & \node[circ2] {}; & \node[circ] {};\\
\node[circ] {}; & \node[circ2] {}; &\node[] {\textbf{1}};& \node[circ2] {}; & \node[circ] {};&\node[circ2] {};\\
\node[circ2] {}; &\node[circ] {}; & \node[circ2] {}; &\node[] {\textbf{1}};& \node[circ2] {}; & \node[circ] {};\\
\node[circ] {};& \node[circ2] {};& \node[circ] {}; &\node[circ2] {}; &\node[] {\textbf{1}};& \node[circ2] {};\\
\node[circ2] {}; & \node[circ] {}; & \node[circ2] {}; & \node[circ] {}; & \node[circ2] {}; & \node[] {\textbf{1}};\\
};
\begin{scope}[shift={(11,0)}]
\draw[shift={(0.5,0.5)}] (0,0) grid (6,6);
\node[below left] at (0.5,4) {\normalsize{$P^2_6 (H) =$}};
\matrix[row sep=-0.5mm,column sep=-0.2mm,shift={(1.75,1.75)}] {
&\node[] {\textbf{1}}; &\node[circ2] {}; &\node[circ] {}; &\node[circ2] {}; & \node[circ] {}; \\
\node[] {\textbf{1}}; && \node[] {\textbf{1}};  &\node[circ2] {}; & \node[circ] {}; &\node[circ2] {}; \\
\node[circ2] {}; & \node[] {\textbf{1}};  && \node[] {\textbf{1}};  &\node[circ2] {}; &\node[circ] {}; \\
\node[circ] {}; &\node[circ2] {}; & \node[] {\textbf{1}};  && \node[] {\textbf{1}};  & \node[circ2] {}; \\
\node[circ2] {}; &\node[circ] {}; &\node[circ2] {}; & \node[] {\textbf{1}};  &&\node[] {\textbf{1}}; \\
\node[circ] {}; & \node[circ2] {}; &\node[circ] {}; &\node[circ2] {}; &\node[] {\textbf{1}}; &\\
};
\end{scope}
\begin{scope}[shift={(22,0)}]
\draw[shift={(0.5,0.5)}] (0,0) grid (6,6);
\node[below left] at (0.5,4) {\normalsize{$P^3_6 (H) =$}};
\matrix[row sep=-0.5mm,column sep=-0.2mm,shift={(1.75,1.75)}] {
&&\node[] {\textbf{1}}; &\node[circ2] {}; &\node[circ] {}; & \node[circ2] {}; \\
&\node[] {\textbf{1}}; &\node[circ2] {}; &\node[circ] {}; & \node[circ2] {}; &\node[circ] {}; \\
\node[] {\textbf{1}}; &\node[circ2] {}; &\node[circ] {}; & \node[circ2] {};&\node[circ] {}; &\node[circ2] {}; \\
\node[circ2] {}; &\node[circ] {}; &\node[circ2] {}; &\node[circ] {}; &\node[circ2] {}; & \node[] {\textbf{1}}; \\
\node[circ] {}; &\node[circ2] {}; &\node[circ] {}; &\node[circ2] {}; &\node[] {\textbf{1}}; &\\
\node[circ2] {}; & \node[circ] {}; &\node[circ2] {}; &\node[] {\textbf{1}}; &&\\
};
\node[below left] at (2.5,3.5) {\textbf{1}};
\node[below left] at (3.5,2.5) {\textbf{1}};
\node[below left] at (3.5,4.5) {\textbf{1}};
\node[below left] at (4.5,3.5) {\textbf{1}};
\node[below left] at (4.5,5.5) {\textbf{1}};
\node[below left] at (5.5,4.5){\textbf{1}};
\end{scope}
\begin{scope}[shift={(0,-8)}]
\draw[shift={(0.5,0.5)}] (0,0) grid (6,6);
\node[below left] at (0.5,4) {\normalsize{$P^4_6 (H) =$}};
\matrix[row sep=-0.5mm,column sep=-0.2mm,shift={(1.75,1.75)}] {
&&&\node[] {\textbf{1}}; &\node[circ2] {}; & \node[circ] {}; \\
&&\node[] {\textbf{1}}; &\node[circ2] {}; & \node[circ] {}; &\node[circ2] {}; \\
&\node[] {\textbf{1}}; &\node[circ2] {}; &\node[circ] {}; &\node[circ2] {}; &\node[] {\textbf{1}}; \\
\node[] {\textbf{1}}; &\node[circ2] {}; &\node[circ] {}; &\node[circ2] {}; &\node[] {\textbf{1}}; & \\
\node[circ2] {}; &\node[circ] {}; &\node[circ2] {}; &\node[] {\textbf{1}}; &&\\
\node[circ] {}; & \node[circ2] {}; &\node[] {\textbf{1}}; &&&\\
};
\node[below left] at (2.5,2.5) {\textbf{1}};
\node[below left] at (3.5,3.5) {\textbf{1}};
\node[below left] at (4.5,4.5) {\textbf{1}};
\node[below left] at (5.5,5.5) {\textbf{1}};
\end{scope}
\begin{scope}[shift={(11,-8)}]
\draw[shift={(0.5,0.5)}] (0,0) grid (6,6);
\node[below left] at (0.5,4) {\normalsize{$P^5_6 (H) =$}};
\matrix[row sep=-0.5mm,column sep=-0.2mm,shift={(1.75,1.75)}] {
&&&&\node[] {\textbf{1}}; & \node[circ2] {}; \\
&&&\node[] {\textbf{1}}; & \node[circ2] {}; &\node[] {\textbf{1}}; \\
&&\node[] {\textbf{1}}; & \node[circ2] {};&\node[] {\textbf{1}}; &\\
&\node[] {\textbf{1}}; &\node[circ2] {}; &\node[] {\textbf{1}}; && \\
\node[] {\textbf{1}}; &\node[circ2] {}; &\node[] {\textbf{1}}; &&&\\
\node[circ2] {}; &\node[] {\textbf{1}}; &&&&\\
};
\end{scope}
\begin{scope}[shift={(22,-8)}]
\draw[shift={(0.5,0.5)}] (0,0) grid (6,6);
\node[below left] at (0.5,4) {\normalsize{$P^6_6 (H) =$}};
\matrix[row sep=-0.5mm,column sep=-0.2mm,shift={(1.75,1.75)}] {
&&&&&\node[] {\textbf{1}}; \\
&&&&\node[] {\textbf{1}}; &\\
&&&\node[] {\textbf{1}}; &&\\
&&\node[] {\textbf{1}}; &&&\\
&\node[] {\textbf{1}}; &&&&\\
\node[] {\textbf{1}}; &&&&&\\
};
\end{scope}
\end{tikzpicture}
\captionsetup{width=0.85\textwidth}
\caption{\emph{Matrix elements for $n=6$ pseudometrics of the general model of \autoref{ourHamiltonian}. Red dots for imaginary entries and green dots for real entries.}}
\label{generalFig}
\end{figure}

We start our discussion at dimension 4. This is a compromise between the ability to demonstrate extrapolation patterns and ability to admit brute-force construction through symbolic manipulation software (in our case MAPLE). Before delving into full machinery of symbolic manipulations, we remind that the general Hamiltonian may be understood as a (sufficiently small) perturbation of the discrete Laplacian $(\Delta_n)_{ij} = -\delta_{i, i+1} + 2 \delta_{i, i} -\delta_{i+1, i}$, with the diagonal terms being set to zero by appropriate energy shift. It is instructive to recall that the general pseudometric construction for such a (hermitian) operator yields the sequence

\begin{equation}
P^1_4 (\Delta) = 
\begin{bmatrix}
1 &  &  &  \\
 & 1  &  &  \\
 &  & 1 &   \\
 &  &  & 1 \end{bmatrix}
\;\;\;\; P^2_4  (\Delta) = \begin{bmatrix}
 & 1  &  & \\
1 &  & 1 & \\
 & 1 &  & 1 \\
 &  & 1 &  \end{bmatrix} 
\;\;\;\;
P^3_4  (\Delta) = \begin{bmatrix}
& & 1 &  \\
& 1 &  & 1 \\
1 & & 1 &  \\
 & 1 & & \end{bmatrix}  
\;\;\;\;
P^4_4  (\Delta) = \begin{bmatrix}
&&& 1 \\
&&1& \\
&1&& \\
1&&& \end{bmatrix}
\label{LaplacianPseudo}
\end{equation}

with the extrapolation pattern for $n >4$ being clear. Since $\Delta$ is hermitian, the identity operator belongs among admissible metrics (and we have chosen our sequence such that the identity appears there explicitly). Also, note that the identity is the only positive matrix in the sequence. On the other side of the sequence, the last pseudometric represents a discrete operator of parity. In our low-dimensional setting, we can repeat the brute-force symbolic manipulation techniques to reveal the sequence of pseudometrics

\begin{align}
\begin{split}
& P^1_4 (H) = 
\begin{bmatrix}
1 & i\alpha & -\alpha (\alpha+\beta) & -i ( {\alpha} (\alpha^2 - \beta^2)-\beta) \\ 
-i\alpha&1&i \left( \alpha+\beta \right) &-\alpha \left( \alpha+\beta \right) \\ 
-\alpha\, \left( \alpha+\beta \right) &-i \left( \alpha+\beta \right) &1&i\alpha \\ 
i ( {\alpha} (\alpha^2 - \beta^2)-\beta) &-\alpha ( \alpha+\beta ) &-i\alpha&1
\end{bmatrix} \\
&  \;\;\qquad \qquad P^2_4 (H) = \begin{bmatrix}
 & 1  & i (\alpha + \beta) & \beta^2 - \alpha^2 \\
1 &   & 1 & i (\alpha + \beta) \\
-i (\alpha + \beta) & 1 &  & 1 \\
\beta^2 - \alpha^2 & -i (\alpha + \beta) & 1 &  \end{bmatrix}  \\
&\;\;\; \qquad P^3_4 (H) = \begin{bmatrix}
&& 1 & \mathrm{i} \alpha \\
& 1 & \mathrm{i} \beta & 1 \\
1 & - \mathrm{i} \beta & 1 &  \\
- \mathrm{i} \alpha & 1 && \end{bmatrix}  
\qquad \qquad \qquad 
P^4_4 (H) = \begin{bmatrix}
&&& 1 \\
&&1& \\
&1&& \\
1&&& \end{bmatrix}
\label{general4}
\end{split}
\end{align}

which clearly demonstrates the preservation of structure of \autoref{LaplacianPseudo}. The occurence of the last pseudometric of the sequence in an unchanged form is a direct consequence of operator $\mathscr{PT}$-symmetry. In all cases discussed here, we shall be constructing the pseudometrics with the same index labeling, where the $k$-th pseudometric is a parameter-dependent perturbation of the $k$-th pseudometric for the discrete Laplacian. This pattern may be further verified in higher dimensions. It is expressed schematically for $n=6$ in \autoref{generalFig}.

Although the general form of matrix elements is too complicated to be expressed explicitly with growing $n$, we may still employ a useful ansatz for the pseudometrics, which has a unified form for any $n \in \mathbb{N}$. The ansatz claims, that the nontrivial matrix elements occupy only a finite banded part of the matrix aligning along its antidiagonals. In the schematic drawing of \autoref{generalFig}, we express this ansatz and its trivial, real and purely complex entries for the pseudometrics with, respectively, ones, red dots and green dots (as before, the extrapolating pattern is clear enough).

\section{The special cases}

In this section, we feel motivated by the possibility of simplification of the pseudometric formulae by appropriate choice of parameters. We do not claim to exhaust all the interesting special cases, but merely point our at some interesting subclasses of \autoref{ourHamiltonian}, which have not yet received sufficient attention in the literature. Our first approach is directly inspired by \cite{myMyMy}, where, motivated by recent results in physics of condensed matter, a nicely extrapolating family of pseudometrics was constructed for

\begin{equation}
\alpha = \alpha \qquad \beta = \gamma = \dots = 0
\label{myRobin}
\end{equation}

\begin{figure}[h!]
\centering
\begin{tikzpicture}[circ/.style = {circle,radius=2mm,fill=green},circ2/.style = {circle,radius=2mm,fill=red},scale=0.5]
\draw[shift={(0.5,0.5)}] (0,0) grid (6,6);
\node[below left] at (0.5,4) {\normalsize{$P^1_6 (H_\gamma) =$}};
\matrix[row sep=-0.5mm,column sep=-0.2mm,shift={(1.75,1.75)}] {
\node[] {\textbf{1}}; &&&&&\node[circ2] {}; \\
&\node[] {\textbf{1}};&&&\node[circ2] {}; &\\
&&\node[] {\textbf{1}};& \node[circ2] {}; &&\\
&& \node[circ2] {}; &\node[] {\textbf{1}};&&\\
&\node[circ2] {}; &&&\node[] {\textbf{1}};&\\
\node[circ2] {}; &&&&& \node[] {\textbf{1}};\\
};
\begin{scope}[shift={(11,0)}]
\draw[shift={(0.5,0.5)}] (0,0) grid (6,6);
\node[below left] at (0.5,4) {\normalsize{$P^2_6 (H_\gamma) =$}};
\matrix[row sep=-0.5mm,column sep=-0.2mm,shift={(1.75,1.75)}] {
&\node[] {\textbf{1}}; &&&\node[circ2] {};& \\
\node[] {\textbf{1}}; &&\node[] {\textbf{1}}; &\node[circ2] {};&&\node[circ2] {};\\
&\node[] {\textbf{1}}; && \node[circ] {};&\node[circ2] {};&\\
&\node[circ2] {};&\node[circ] {};&&\node[] {\textbf{1}}; & \\
\node[circ2] {};&&\node[circ2] {};&\node[] {\textbf{1}}; &&\node[] {\textbf{1}}; \\
&\node[circ2] {};&&&\node[] {\textbf{1}}; &\\
};
\node[below left] at (4.5,4.5) {\textbf{1}};
\node[below left] at (3.5,3.5){\textbf{1}};
\end{scope}
\begin{scope}[shift={(22,0)}]
\draw[shift={(0.5,0.5)}] (0,0) grid (6,6);
\node[below left] at (0.5,4) {\normalsize{$P^3_6 (H_\gamma) =$}};
\matrix[row sep=-0.5mm,column sep=-0.2mm,shift={(1.75,1.75)}] {
&&\node[] {\textbf{1}}; &\node[circ2] {}; &\node[circ] {};&\\
&\node[] {\textbf{1}}; &&\node[] {\textbf{1}}; &\node[circ2] {};&\node[circ] {};\\
\node[] {\textbf{1}}; &&\node[] {\textbf{1}}; &\node[circ2] {};&\node[] {\textbf{1}}; &\node[circ2] {}; \\
\node[circ2] {}; &\node[] {\textbf{1}}; &\node[circ2] {};&\node[] {\textbf{1}}; && \node[] {\textbf{1}}; \\
\node[circ] {}; &\node[circ2] {};&\node[] {\textbf{1}}; &&\node[] {\textbf{1}}; &\\
&\node[circ] {}; &\node[circ2] {}; &\node[] {\textbf{1}}; &&\\
};
\end{scope}
\begin{scope}[shift={(0,-8)}]
\draw[shift={(0.5,0.5)}] (0,0) grid (6,6);
\node[below left] at (0.5,4) {\normalsize{$P^4_6 (H_\gamma) =$}};
\matrix[row sep=-0.5mm,column sep=-0.2mm,shift={(1.75,1.75)}] {
&&&\node[] {\textbf{1}}; && \node[circ] {};\\
&&\node[] {\textbf{1}}; &\node[circ2] {};& \node[circ] {};&\\
&\node[] {\textbf{1}}; && \node[] {\textbf{1}}; &\node[circ2] {};&\node[] {\textbf{1}}; \\
\node[] {\textbf{1}}; &\node[circ2] {}; &\node[] {\textbf{1}};&&\node[] {\textbf{1}}; & \\
& \node[circ] {}; &\node[circ2] {}; &\node[] {\textbf{1}}; &&\\
\node[circ] {};&&\node[] {\textbf{1}}; &&&\\
};
\node[below left] at (2.5,2.5) {\textbf{1}};
\node[below left] at (5.5,5.5) {\textbf{1}};
\end{scope}
\begin{scope}[shift={(11,-8)}]
\draw[shift={(0.5,0.5)}] (0,0) grid (6,6);
\node[below left] at (0.5,4) {\normalsize{$P^5_6 (H_\gamma) =$}};
\matrix[row sep=-0.5mm,column sep=-0.2mm,shift={(1.75,1.75)}] {
&&&&\node[] {\textbf{1}}; &\\
&&&\node[] {\textbf{1}}; &&\node[] {\textbf{1}}; \\
&&\node[] {\textbf{1}}; & \node[circ2] {};&\node[] {\textbf{1}}; &\\
&\node[] {\textbf{1}}; &\node[circ2] {}; &\node[] {\textbf{1}}; && \\
\node[] {\textbf{1}}; &&\node[] {\textbf{1}}; &&&\\
&\node[] {\textbf{1}}; &&&&\\
};
\end{scope}
\begin{scope}[shift={(22,-8)}]
\draw[shift={(0.5,0.5)}] (0,0) grid (6,6);
\node[below left] at (0.5,4) {\normalsize{$P^6_6 (H_\gamma) =$}};
\matrix[row sep=-0.5mm,column sep=-0.2mm,shift={(1.75,1.75)}] {
&&&&&\node[] {\textbf{1}}; \\
&&&&\node[] {\textbf{1}}; &\\
&&&\node[] {\textbf{1}}; &&\\
&&\node[] {\textbf{1}}; &&&\\
&\node[] {\textbf{1}}; &&&&\\
\node[] {\textbf{1}}; &&&&&\\
};
\end{scope}
\end{tikzpicture}
\captionsetup{width=0.85\textwidth}
\caption{\emph{Entries for $n=6$ pseudometrics for $\alpha = \beta = 0$ with red dots standing for $\pm \mathrm{i} \gamma$ and green dots for $(i \gamma)^2$}}
\label{almostFinal}
\end{figure}
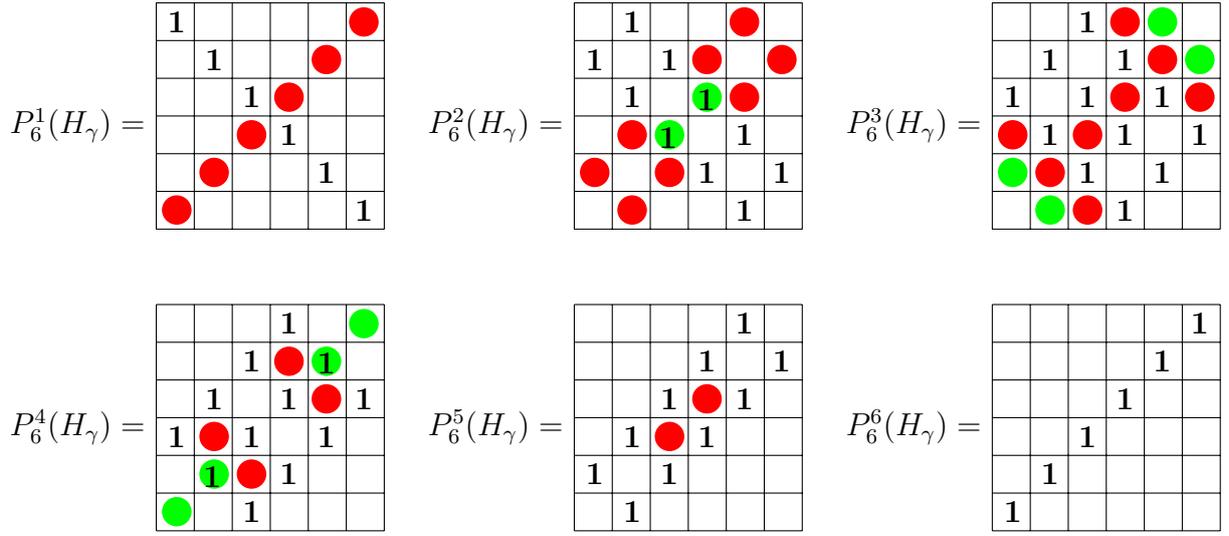

This particular case, while not diminishing the sparsity structure of the pseudometrics in any substantial way, deserves particular attention in the uncommon elegance and expressibility of the resulting formulas. Here, in parallel with the numerical study in \autoref{dobs1D}, we aim to extend this consideration to arbitrary single non-zero parameter. One of the motivations for such a treatment might be the discretization origin of the model of \autoref{myRobin} in \cite{discreteRobin}, which opens the possibility, that other point-interaction differential operators correspond to these generalized models. Of course, the case of the $n$-th parameter being nonzero can be studied only the matrices of size $n \times n$ and larger. We restate the case $n=4$ for $\alpha = 0$, which may form the basis of further development

\begin{align}
\begin{split}
& P^1_4 (H_\beta) = 
\begin{bmatrix}
1 &  &  & \mathrm{i} \beta \\ 
&1&i \beta & \\ 
&-i \beta &1& \\ 
- \mathrm{i} \beta &&&1
\end{bmatrix}  \qquad
P^2_4 (H_\beta) = \begin{bmatrix}
 & 1  & i \beta & - (i \beta)^2  \\
1 &   & 1 & i \beta \\
- i \beta & 1 &  & 1 \\
- (i \beta)^2  & - \mathrm{i} \beta & 1 &  \end{bmatrix}  \\ 
& \qquad \;\;\; P^3_4 (H_\beta) = \begin{bmatrix}
&& 1 &  \\
& 1 & \mathrm{i} \beta & 1 \\
1 & - \mathrm{i} \beta & 1 &  \\
 & 1 && \end{bmatrix}  \qquad 
P^4_4 (H_\beta) = \begin{bmatrix}
&&& 1 \\
&&1& \\
&1&& \\
1&&& \end{bmatrix}
\end{split}
\end{align}

which shows a further progression in terms of cancellation of another matrix elements. Indeed, evidence suggests that each general choice of parameter $p$ suggests to possess a nicely extrapolating family of pseudometrics. For taste of the schemes, we provide yet the pseudometric for the parameter $\gamma$ and $n=6$. They are shown in \autoref{almostFinal}

So far, all the single-site interactions seem to admit formulas sufficiently simple, so that the complexity of the resulting matrix elements does not grow beyond the expressibility by useful closed formulas. The full treatment of this case is, however, beyond the scope of the present article, mainly because for larger parameter values one needs to begin consideration at matrices of large dimensions, which are not suitable for explicit plotting. Despite this fact, the message of this treatment is clear: the friendly form of the general pseudometrics is not restricted to the boundary interaction.

On the other side of the single-site interaction in the parametric spectrum lies the full-lattice interaction of the model \autoref{ourHamiltonian}. By the full lattice interaction, we understand the case where none of the parameters is zero at any moment. Guided by the desire for the simplification of our pseudometric patterns, and also by the explicit formulas for $n=4$, we are tempted to choose the most friendly and matrix element eliminating interaction in an alternating form

\begin{equation}
\alpha = -\beta = \gamma = - \delta = \dots
\end{equation}

\begin{figure}[h!]
\centering
\begin{tikzpicture}[circ/.style = {circle,radius=2mm,fill=violet},circ2/.style = {circle,radius=2mm,fill=red},scale=0.5]
\draw[shift={(0.5,0.5)}] (0,0) grid (6,6);
\node[below left] at (0.5,4) {\normalsize{$P^1_6 (H_a) =$}};
\matrix[row sep=-0.5mm,column sep=-0.2mm,shift={(1.75,1.75)}] {
\node[] {\textbf{1}}; & \node[circ2] {}; && \node[circ2] {}; &&\node[circ2] {}; \\
\node[circ2] {}; &\node[] {\textbf{1}};&&&&\\
&&\node[] {\textbf{1}};& \node[circ2] {}; &&\node[circ2] {}; \\
\node[circ2] {}; && \node[circ2] {}; &\node[] {\textbf{1}};&&\\
&&&&\node[] {\textbf{1}};& \node[circ2] {};\\
\node[circ2] {}; &&\node[circ2] {}; && \node[circ2] {}; & \node[] {\textbf{1}};\\
};
\begin{scope}[shift={(11,0)}]
\draw[shift={(0.5,0.5)}] (0,0) grid (6,6);
\node[below left] at (0.5,4) {\normalsize{$P^2_6 (H_a) =$}};
\matrix[row sep=-0.5mm,column sep=-0.2mm,shift={(1.75,1.75)}] {
&\node[] {\textbf{1}}; &&&& \\
\node[] {\textbf{1}}; &&\node[] {\textbf{1}}; &&&\\
&\node[] {\textbf{1}}; && \node[] {\textbf{1}}; &&\\
&&\node[] {\textbf{1}};&&\node[] {\textbf{1}}; & \\
&&&\node[] {\textbf{1}}; &&\node[] {\textbf{1}}; \\
&&&&\node[] {\textbf{1}}; &\\
};
\end{scope}
\begin{scope}[shift={(22,0)}]
\draw[shift={(0.5,0.5)}] (0,0) grid (6,6);
\node[below left] at (0.5,4) {\normalsize{$P^3_6 (H_a) =$}};
\matrix[row sep=-0.5mm,column sep=-0.2mm,shift={(1.75,1.75)}] {
&&\node[] {\textbf{1}}; &\node[circ2] {}; && \node[circ2] {}; \\
&\node[] {\textbf{1}}; &\node[circ2] {}; &\node[] {\textbf{1}}; &&\\
\node[] {\textbf{1}}; &\node[circ2] {}; &\node[] {\textbf{1}}; &&\node[] {\textbf{1}}; &\node[circ2] {}; \\
\node[circ2] {}; &\node[] {\textbf{1}}; &&\node[] {\textbf{1}}; &\node[circ2] {}; & \node[] {\textbf{1}}; \\
&&\node[] {\textbf{1}}; &\node[circ2] {}; &\node[] {\textbf{1}}; &\\
\node[circ2] {}; &&\node[circ2] {}; &\node[] {\textbf{1}}; &&\\
};
\end{scope}
\begin{scope}[shift={(0,-8)}]
\draw[shift={(0.5,0.5)}] (0,0) grid (6,6);
\node[below left] at (0.5,4) {\normalsize{$P^4_6 (H_a) =$}};
\matrix[row sep=-0.5mm,column sep=-0.2mm,shift={(1.75,1.75)}] {
&&&\node[] {\textbf{1}}; && \\
&&\node[] {\textbf{1}}; && \node[] {\textbf{1}}; &\\
&\node[] {\textbf{1}}; && \node[] {\textbf{1}}; &&\node[] {\textbf{1}}; \\
\node[] {\textbf{1}}; &&\node[] {\textbf{1}};&&\node[] {\textbf{1}}; & \\
& \node[] {\textbf{1}}; &&\node[] {\textbf{1}}; &&\\
&&\node[] {\textbf{1}}; &&&\\
};
\end{scope}
\begin{scope}[shift={(11,-8)}]
\draw[shift={(0.5,0.5)}] (0,0) grid (6,6);
\node[below left] at (0.5,4) {\normalsize{$P^5_6 (H_a) =$}};
\matrix[row sep=-0.5mm,column sep=-0.2mm,shift={(1.75,1.75)}] {
&&&&\node[] {\textbf{1}}; & \node[circ2] {}; \\
&&&\node[] {\textbf{1}}; & \node[circ2] {}; &\node[] {\textbf{1}}; \\
&&\node[] {\textbf{1}}; & \node[circ2] {};&\node[] {\textbf{1}}; &\\
&\node[] {\textbf{1}}; &\node[circ2] {}; &\node[] {\textbf{1}}; && \\
\node[] {\textbf{1}}; &\node[circ2] {}; &\node[] {\textbf{1}}; &&&\\
\node[circ2] {}; &\node[] {\textbf{1}}; &&&&\\
};
\end{scope}
\begin{scope}[shift={(22,-8)}]
\draw[shift={(0.5,0.5)}] (0,0) grid (6,6);
\node[below left] at (0.5,4) {\normalsize{$P^6_6 (H_a) =$}};
\matrix[row sep=-0.5mm,column sep=-0.2mm,shift={(1.75,1.75)}] {
&&&&&\node[] {\textbf{1}}; \\
&&&&\node[] {\textbf{1}}; &\\
&&&\node[] {\textbf{1}}; &&\\
&&\node[] {\textbf{1}}; &&&\\
&\node[] {\textbf{1}}; &&&&\\
\node[] {\textbf{1}}; &&&&&\\
};
\end{scope}
\end{tikzpicture}
\captionsetup{width=0.85\textwidth}
\caption{\emph{Matrix entries for $n=6$ and the alternating lattice interaction, red dots standing for $\pm \mathrm{i} \alpha$.}}
\label{finalFinal}
\end{figure}
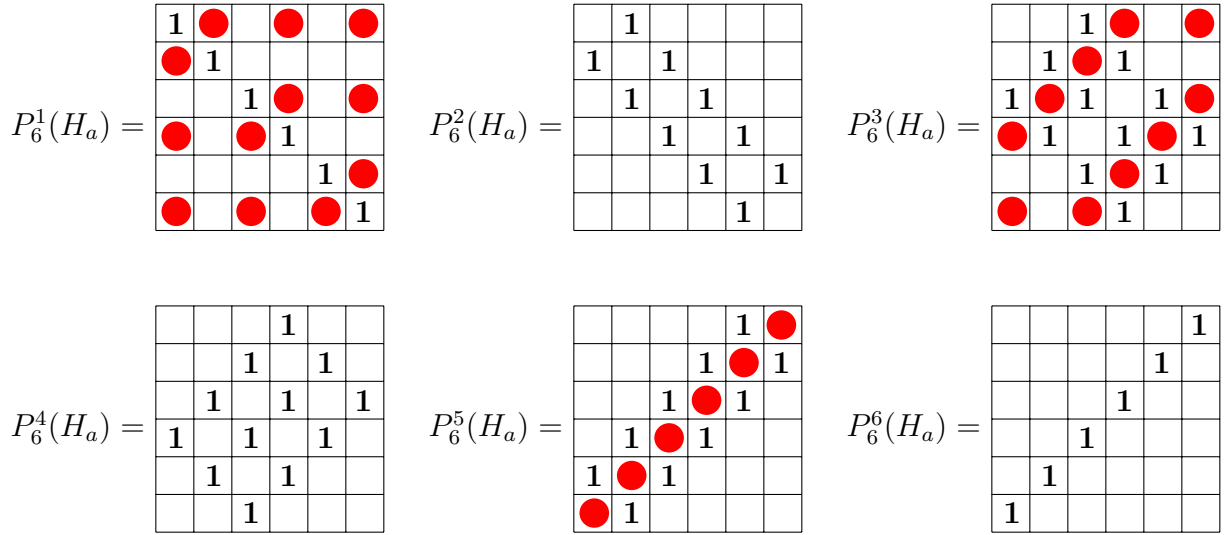

This alternating interaction Hamiltonian shall be denoted as $H_a$. Already the pattern for $n=4$ indicates the extremely simple form of the resulting pseudometrics. The simplicity is demonstrated in the fact, that only one type of nontrivial matrix element exists in whole family of pseudometrics, with the value $\mathrm{i} \alpha$. Consequently, we may drop two-color notation from above, and denote by a red dot simply the explicitly expressible element $\mathrm{i} \alpha$. The results of these considerations are summarized in \autoref{finalFinal}

with the extrapolation pattern again being clear for any dimension. Having succesfully constructed all the pseudometrics, the final part of the task would, in principle, consist of verifying the condition of positivity of the resulting most general $n$-parametric linear combination. As usual in these cases, we have constructed one of the pseudometric positive for sufficiently small values of $\alpha$. Using this fact, we might employ the powerful machinery of perturbation theory and write

\begin{equation}
\Theta^{}_n (H^{}_\alpha) = P^1_n (H^{}_\alpha) + \varepsilon^{}_2 P^2_n (H^{}_\alpha)+ \dots + \varepsilon^{}_n P^n_n (H^{}_\alpha)
\end{equation}

The exceptionally friendly character of our choice of interaction may be seen in the complete triviality of the even-numbered pseudometrics in each sequence, and also in the occurence of a single form of nontrivial matrix element. We have reached the goal we were looking for: we have succesfully constructed a complete set of pseudometrics for the special case of \autoref{general4}, which surpasses the previously examined toy models (e.g. \cite{SSH}) both in sparsity and simplicity of the resulting pseudometric matrix elements.

\section{Discussion}

The current active research in quasi-hermitian representation of quantum mechanics can be roughly divided into three directions, nicely summarized in a recent collection of proceedings \cite{ZnojilBook}. The first one has the goal to clarify the fundamental phenomenological foundations of such a representation, the conceptual problems with time-dependent metric operators, its time evolution or the Heisenberg representation \cite{threeHilbertSpace,HeisenbergRepresentation}. Another direction lies in giving proper mathematical foundation to quasi-hermitian Hamiltonians with unbounded and/or singular metrics \cite{MostafazadehUnbounded,BagarelloUnbounded,KuzhelUnbounded,AntoineUnbounded}. Finally, the third one consists of finding proper solvable quasi-hermitian models, which are what makes quasi-hermitian quantum mechanics a useful concept.

In this paper, we pursued the third direction with focus on finite-dimensional quantum Hamiltonians. Regarding model-building schemes for finite quasi-hermitian Hamiltonians, different techniques may be employed to yield different useful results. The correspondence between useful models of solid state physics and lattice quasi-hermitian operators has proven fruitful very recently \cite{SSH, AubryAndre}. Also, the correspondence between (non-normalized) orthogonal polynomials \cite{SimonOrthopoly} and quantum-mechanical matrix models in has produces some interesting output \cite{ZnojilLaguerre,ZnojilChebyshev}. The nice feature in these particular models is, that the pseudometric may be constructed through a recurrent formula, which can be very often solved explicitly. The third recipe to construct finite-dimensional models has been seen to lie in discretization of infinite-dimensional models, as in \cite{discreteRobin}.

The choice of our Hamiltonian stems from the attempt to provide a general treatment to various finite-dimensional operators with complex interactions scattered in the literature. Our results may be briefly summarized as follows: we addressed the general Hamiltonians for low dimensions symbolically, and found an emergent pseudometric pattern, which was then reconfirmed by further sample calculations. In order to achieve a true simplicity of the pseudometrics, we have carefully selected a number of special cases, again guided by the low-dimensional results. The simplicity of the resulting pseudometrics, in combination with the scope of the interaction not being limited to boundary terms, provides a fresh scheme into presently known quasi-hermitian matrices. It may seem encouraging that such a recipe leaves the construction non-numerical and offers unexpectedly transparent benchmark results. The full-lattice interaction may also have an interesting counterpart in the infinite-dimensional discrete limit (which can be, unlike the boundary interaction, defined with no obstacles), and finally in their appropriately-defined continuous counterparts $V(x)$. To this end, we complement the numerical experiment of section 2 with its counterpart for full-lattice interactions. 

\begin{table}[h!]
\centering
\renewcommand{\arraystretch}{1}
\begin{tabular}{|l||c|c|c|c|}
  \hline
  & $n=10$ & $n=30$ & $n=50$ & $n=100$\\
  \hline \hline
   $\alpha = - \beta = \gamma = - \delta =  \dots$ & 0.2934 & 0.1015 & 0.0623 & 0.0316 \\
  \hline
   $\alpha = \beta = \gamma = \delta =  \dots$ & 0.1413 & 0.0185 & 0.0073 & 0.0018 \\
  \hline
\end{tabular}
\renewcommand{\arraystretch}{1}
\captionsetup{width=0.85\textwidth}
    \caption{\emph{Critical values (exceptional points) for $2$ different full-lattice Hamiltonians of varying dimension $n$.}}
    \label{dobsFull}
\end{table}

Note that the domains of observability are in orders of magnitude smaller that for single-site interactions, and have a clearly defined zero limit as $n \rightarrow \infty$. The possibilities of generalization are vast, the one which looks very promising lies in further systematic search for exactly solvable finite-dimensional models, which would admit not only closed-form representation of the eigenenergies, but also the explicit construction of the pseudometrics. The examination of full-lattice interactions not being localized at the boundaries looks particularly promising. In this direction, it may be worth studying eighter general complex interactions (instead of purely complex ones), or switching attention to finite-range interactions, which would however most likely produce a number of new obstacles to overcome. In general, the field of solvable quasi-hermitian operators is far from being fully explored, and offers exciting new possibilities even in finite-dimensional Hilbert spaces.

\printbibliography

\end{document}